\documentclass[journal,twoside,web]{ieeecolor}
\usepackage{generic}
\usepackage{cite}
\usepackage{amsmath,amsfonts,amssymb}
\usepackage{url}
\usepackage[dvipsnames]{xcolor}
\usepackage[colorlinks=true,linkcolor=Cerulean,citecolor=Cerulean,urlcolor=Cerulean]{hyperref}
\usepackage{orcidlink}

\begin{document}

\title{LLM-Guided Contextual Action Evaluation for Operational Decisions in Industrial Processes}

\author{
  Youcheng Zong\(^{\orcidlink{0009-0008-6795-8412}}\),~\IEEEmembership{Student~Member,~IEEE},
  Runda Jia\(^{\orcidlink{0000-0002-8586-243X}}\),
  \\ and Dakuo He\(^{\orcidlink{0000-0001-8303-529X}}\)
  \thanks{This work was supported by the Fundamental Research Funds for the Central Universities, China (N26GFZ006). \textit{(Corresponding author: Runda Jia.)}}
  \thanks{Youcheng Zong, Runda Jia, and Dakuo He are with the College of Information Science and Engineering, Northeastern University, Shenyang 110004, China (e-mail: youchengzong@stumail.neu.edu.cn; jiarunda@ise.neu.edu.cn; hedakuo@ise.neu.edu.cn).}
}

\maketitle

\begin{abstract}
  Industrial actor--critic methods usually represent continuous actions as anonymous numerical coordinates. They must therefore learn from limited interactions which process variables each action affects, in which direction, and after what delay.
  Fixed industrial documents already describe part of these relations, but their open-text statements neither represent the current operating condition nor directly fit a numerical policy.
  This article presents LLM-Guided Contextual Action Evaluation for Operational Decisions in Industrial Processes (LCAE), which uses a large language model before training to normalize fixed documents into a frozen action--observation--direction--delay relation basis.
  Recent numerical action--response history then modulates the current strength of each relation, while the evaluated action forms a state-conditioned nonlinear action-effect field in the same basis.
  The critic evaluates actions through this field, and the actor uses the same relation gains to generate actions, making document semantics part of maximum-entropy policy learning.
  Neither the LLM nor the embedding model runs online during training or deployment; the deployed policy uses only frozen semantic artifacts and visible numerical history.
  The method states a falsifiable hypothesis: when documented relations are correct and recent history reflects their contextual strength, this action representation should provide a more useful decision bias than raw action coordinates.
\end{abstract}


\section{Introduction}\label{sec:introduction}

Operational decision-making in process industries adjusts manipulated variables from continuous observations and improves long-term operating objectives through delayed process responses.
These tasks are usually partially observable because available measurements provide an incomplete view of the internal process state, and an action may affect several later sampling instants~\cite{kaelbling1998pomdp}.
Reinforcement learning provides a common interface for such closed-loop sequential decisions, and deep actor--critic methods have been used for continuous control and industrial process control~\cite{lillicrap2016ddpg,haarnoja2018sac,nian2020review}.
These studies show that numerical policies can learn complex control laws from interaction, but their action interfaces still usually treat each manipulated variable as an anonymous coordinate.

Anonymous action coordinates leave the critic with a compound learning problem.
To estimate action value, the critic must learn how the state evolves and rediscover which observations each action affects, in which direction, and after what delay.
Twin critics, target networks, and maximum-entropy objectives can improve numerical optimization~\cite{fujimoto2018td3,haarnoja2018sac}, but they do not explain the process meaning of an action coordinate.
In industrial environments with costly interaction, coupled actions, or slow responses, rediscovering these relations consumes samples that could otherwise be used to learn the policy.

Industrial numerical modeling has extracted state and prediction information from numerical trajectories~\cite{zong2023iron,zong2023judgment,zong2025hybridgrid,zong2025metacontrastive}, but industrial systems also contain information outside those trajectories.
Equipment descriptions, variable tables, operating procedures, and environment descriptions often state which visible variables an action affects through process paths, the typical direction of the effect, and the delay range in which a response may occur.
These relations may span several paragraphs, use inconsistent terminology, and mix structured fields with open text.
An LLM can use open knowledge and language reasoning to normalize these descriptions into consistent relations, but the documents themselves remain static.
A document can describe a basic effect path, but it cannot state how strong that path is in the current operating condition or directly specify the optimal action.

Variable-name matching or fixed rules alone are often insufficient for this conversion.
The same manipulated variable may appear as an equipment tag, a control-loop name, or a process term, while direction and delay statements may be distributed across different document locations.
The necessary role of the LLM in this work is to resolve such open vocabulary and cross-sentence relations and then restrict the result to fields allowed by the task schema.
This role differs from replacing the controller with a language model: the LLM constructs the relation interface, while the numerical policy still learns current decisions from interaction.

This distinction determines how this work separates language and numerical information.
Fixed documents define action-relation coordinates that are expected to remain stable across operating conditions, while recent action--response history estimates their relative strength in the current state.
The evaluated action must also retain its own amplitude because different actions cannot have the same value under identical relation strengths.
The required action representation must therefore depend jointly on fixed relation identity, recent numerical context, and the current candidate action.

This article presents LLM-Guided Contextual Action Evaluation (LCAE) to implement this division of roles.
Before training, an LLM reads only fixed environment documents and a task schema and produces action--observation--direction--delay relation cards.
A fixed renderer and embedding model convert the cards into a frozen semantic basis, while the relation fields determine action selection, observation selection, and valid delay bands.
An online numerical module extracts relation-wise action--response events from recent history and computes relation gains centered at one.
The current action and these gains form a state-conditioned action-effect field in the frozen relation basis. The critic uses this field to estimate long-term value, and the actor uses the same gains to generate the action that will be executed.

LCAE does not place an LLM in the control loop.
The LLM does not read numerical trajectories, rewards, hidden simulator states, optimal actions, or test information, and it does not generate, rank, or validate online actions.
The embedding model also runs only before training.
Closed-loop decisions during training and deployment are made entirely by a compact numerical actor, separating semantic construction cost from per-step control latency.

This work focuses on a scientific hypothesis rather than presuming a performance improvement.
If fixed documents provide stable action paths and recent history contains their contextual strength, LCAE should improve policy learning and closed-loop return under limited interaction or changing action effectiveness.
Conversely, if a matched-budget raw-action baseline reaches the same performance, or shuffling action--effect correspondences does not change the result, then the documented relations do not provide the expected decision bias.
This falsifiable boundary limits the claim to contextual action evaluation rather than causal-effect identification, fault diagnosis, or safety control.

The main contributions are as follows.
\begin{itemize}
  \item
        We reformulate action learning in industrial actor--critic as contextual action evaluation: fixed documents define relation identity, recent numerical history defines current relation strength, and the candidate action defines relation amplitude.
  \item
        We introduce a state-conditioned action-effect field that combines relation semantics, historical gains, and continuous actions through a bounded nonlinearity, with explicit empty-history fallback, local sensitivity, and action-coverage conditions.
  \item
        We give a complete maximum-entropy actor--critic interface, training objective, and deployment boundary so that frozen offline semantic relations organize both action generation and value learning without an online LLM.
\end{itemize}

The remainder of this article is organized as follows.
Section~\ref{sec:related_work} reviews action representation, language-informed reinforcement learning, and knowledge-guided industrial control.
Section~\ref{sec:method} defines LCAE, and Section~\ref{sec:conclusion} summarizes its claims and boundaries.

\section{Related Work}\label{sec:related_work}

\subsection{From Continuous Control to Industrial Actor--Critic}

Reinforcement learning formulates control as maximizing long-term return through environment interaction~\cite{sutton2018rl}.
DDPG extends deterministic policy gradients to deep continuous control, TD3 reduces function-approximation error through twin critics and delayed updates, and SAC uses a stochastic policy and maximum-entropy objective to improve exploration and training stability~\cite{lillicrap2016ddpg,fujimoto2018td3,haarnoja2018sac}.
Industrial studies have applied deep reinforcement learning to process control and discussed sample efficiency, constraints, stability, and deployment risk~\cite{spielberg2019selfdriving,nian2020review}.
A continuous polymerization case further shows that a numerical actor can directly generate manipulated actions~\cite{ma2019polymerization}.
These routes mainly improve how policies and value functions are optimized, while the action usually still enters the critic as a raw numerical vector.
LCAE builds on this interface: it retains maximum-entropy actor--critic learning but changes the action representation received by the critic and the relation context visible to the actor.
This position also separates the roles of numerical state and action representations.
The history encoder must still summarize load, inventory, setpoints, and other operating conditions, while the action-effect field specifically determines how a candidate action should enter the value function under that history.
LCAE therefore does not assume that semantic relations replace dynamic state modeling, and it does not change the environment's reward or transition definition.

\subsection{From Action Identity to Action Relations}

Action-representation studies have shown that structure among actions can improve generalization across actions.
Act2Vec extracts action relations from demonstration contexts and uses the learned vectors for state augmentation and value-function approximation~\cite{tennenholtz2019actions}.
Another approach jointly learns a low-dimensional action representation and its mapping to actual actions, allowing outcomes of similar actions to be reused in large finite action sets~\cite{chandak2019actionrep}.
Know Your Action Set explicitly learns relations in variable action sets so that a policy can handle action combinations that were not jointly available during training~\cite{jain2022actionset}.
Action-adaptive policies can also infer action impact from recent state changes and adjust when actions are missing or their effects are perturbed~\cite{zeng2023actionimpact}.
These studies learn action structure from demonstrations, action sets, or online state changes, but they do not combine action--observation--direction--delay relations from fixed industrial documents with the current numerical amplitude of a continuous action.
LCAE retains both stable relation identity from documents and contextual strength from history and evaluates continuous actions in the same effect field.
Compared with discrete action embeddings, continuous industrial actions also require amplitude and sign to be preserved.
Mapping an increase and a decrease in valve opening to one static action vector would lose control meaning, while learning only local action impact would discard cross-variable paths stated in documents.
The relation-wise nonlinearity in LCAE preserves both elements and allows one action coordinate to correspond to several observation paths and delay bands.

\subsection{From Language Knowledge to Decision Interfaces}

Natural language can provide goals, policy advice, reward structure, and environment knowledge to reinforcement learning, and these interfaces have been summarized in prior work~\cite{luketina2019language}.
EMMA grounds entities and dynamics from free-form manuals to observations to support policy generalization to unseen tasks~\cite{hanjie2021grounding}.
RLang uses a formal declarative language to describe partial knowledge about all components of an MDP and grounds it into an algorithm-independent partial world model and policy~\cite{rodriguez2023rlang}.
Reward machines represent task-reward structure as automata that reinforcement-learning algorithms can exploit~\cite{icarte2022rewardmachines}.
These methods show that language or symbolic knowledge can change the information available to decision learning rather than serving only as a prediction label.

Foundation models can also transfer pretrained representations to zero-shot domain tasks~\cite{su2026zero}, while LLMs further provide the ability to propose goals or construct policy priors from open descriptions.
ELLM rewards pretraining behaviors with LLM-suggested goals, while GLAM updates an LLM policy through online reinforcement learning~\cite{du2023pretraining,carta2023grounding}.
Both routes place the language model closer to the online decision process.
LCAE extends the offline semantic-interface principle from industrial forecasting~\cite{zong2026tsf} to action evaluation: the LLM only normalizes fixed industrial relations before training, while the online actor and critic remain numerical models.
LCAE therefore studies how language knowledge defines action-evaluation coordinates that numerical history can activate, rather than studying a language policy itself.
The offline interface also allows relation artifacts to be inspected manually or programmatically before training.
Compared with online free-text prompting, this interface can freeze document versions, relation fields, rendered text, and embeddings, ensuring that different training runs receive exactly the same semantic input.
Auditability does not guarantee that a relation is correct, but it makes erroneous relations identifiable and removable rather than hiding them inside per-step language-model calls.

\subsection{From Industrial Knowledge Guidance to Contextual Action Evaluation}

Industrial reinforcement learning has used rules, simulators, and process knowledge to narrow the search space or constrain policies.
Chemical-operation procedure synthesis combines external knowledge, dynamic simulation, and deep reinforcement learning to search for interpretable operating steps~\cite{kubosawa2019procedures}.
Knowledge--data-guided methods use process-causality structure to organize industrial reinforcement learning, and multi-objective graph reinforcement learning uses expert prior graphs to guide fermentation optimization~\cite{zhang2024causality,li2025priorgraph}.
These studies show that industrial knowledge can provide structure beyond numerical interaction, but they usually require manual rules, explicit graphs, or task-specific knowledge encoding.

Another route places LLM agents directly in industrial decisions.
Systems have used LLMs to interpret real-time events, generate production plans, and control automation operations~\cite{xia2025automation}, and other work has used them for human--AI collaborative industrial decision support~\cite{zong2026llmdecision}.
Agent frameworks for fault-tolerant control also generate recovery actions and validate them with digital twins, knowledge graphs, and deterministic constraints~\cite{vyas2026faulttolerant}.
These systems target planning, recovery, or human-decision layers, whose time scales and safety interfaces differ from a closed-loop policy that produces continuous actions with a compact actor at every sampling instant.
LCAE therefore restricts the LLM to offline relation construction and focuses on contextual action evaluation jointly defined by document relations and recent numerical history.
This distinction also limits the deployment claim.
LCAE provides no safety guarantee for LLM-generated actions and does not place language-reasoning latency inside the real-time control budget because the deployed policy never calls the LLM.
Its required engineering checks are offline relation correctness, action-space coverage, and standard validation of the online numerical module.

\section{Method}\label{sec:method}

\subsection{Problem Formulation and Decision Coordinates}\label{sec:problem}

Consider a partially observable industrial decision process in environment $e$.
Let $o_t^{\mathrm{phy}}\in\mathcal O_e^{\mathrm{phy}}$ and $a_t^{\mathrm{phy}}\in\mathcal A_e^{\mathrm{phy}}$ denote the physical observation and physical action, respectively.
Fixed elementwise strictly increasing affine bijections $\mathcal N_e^o$ and $\mathcal N_e^a$ define the internal decision coordinates:
\begin{equation}
  o_t=\mathcal N_e^o\!\left(o_t^{\mathrm{phy}}\right)\in\mathbb R^{d_o},
  \qquad
  a_t=\mathcal N_e^a\!\left(a_t^{\mathrm{phy}}\right)
  \in\mathcal A_e\subseteq\mathbb R^{d_a}.
  \label{eq:coordinates}
\end{equation}
Strict monotonicity preserves the signs of action and response directions stated in the documents.
Except for quantities carrying the $\mathrm{phy}$ superscript at the environment interface, all observations, actions, histories, and policy distributions below use these decision coordinates.

The policy-visible history at decision instant $t$ is
\begin{equation}
  \mathcal H_t=
  \left(o_{t-L:t},a_{t-L:t-1},\Delta t_{t-L:t-1}\right),
  \label{eq:history}
\end{equation}
where $\Delta t_i$ is the physical time elapsed after executing action $a_i$.
The policy generates $a_t$, the environment executes $a_t^{\mathrm{phy}}=(\mathcal N_e^a)^{-1}(a_t)$, and then returns reward $r_t$ and the next observation $o_{t+1}$.
This history definition covers both fixed-step and irregularly timed decision processes.

For a maximum-entropy policy $\pi$, the action value with continuous-time discounting is
\begin{align}
Q^\pi(\mathcal H_t,a_t)
=\mathbb E_\pi\!\Bigg[
&r_t+\sum_{k=1}^{\infty}
\exp\!\left(-\lambda\sum_{j=0}^{k-1}\Delta t_{t+j}\right) \nonumber\\
&\cdot\left(r_{t+k}-\alpha\log\pi(a_{t+k}\mid\mathcal H_{t+k})\right)
\,\Bigm|\,\mathcal H_t,a_t
\Bigg],
\label{eq:q_definition}
\end{align}
where $\lambda>0$ is the temporal discount rate and $\alpha>0$ is the entropy temperature.
When every $\Delta t_t$ is fixed, $\exp(-\lambda\Delta t_t)$ reduces to the standard discrete discount factor $\gamma$.
LCAE retains this decision objective but does not use raw action $a$ directly as the critic's action input.

\subsection{Offline Action-Effect Relation Basis}\label{sec:relation_basis}

For environment $e$, let $D_e$ and $\mathcal S_e$ denote fixed industrial documents and the task schema.
The LLM runs only before policy training and produces a set of relation cards
\begin{equation}
  \mathcal K_e
  =F_{\mathrm{LLM}}(D_e,\mathcal S_e)
  =\{c_p\}_{p=1}^{P_e}.
  \label{eq:cards}
\end{equation}
Each card $c_p$ specifies an action field, an affected visible observation, a typical effect direction, and a physical response-delay band.
A relation card may reference only actions and visible observations declared in the task schema.
It contains no reward, hidden state, future measurement, test information, or recommended action.

Each relation card should also retain its document-evidence location and schema version so that the relation artifact can be traced to fixed inputs.
Structural validation rejects unknown fields, invisible observations, undeclared actions, empty delay bands, and relations outside the allowed scope.
Semantic validation checks whether direction and delay are supported by the recorded document span rather than freely completed by the model.
These checks occur before embedding so that invalid cards cannot enter the frozen relation basis.

A fixed renderer $\mathcal R$ turns each card into canonical text, and a fixed embedding model $E_\psi$ then produces a unit semantic direction:
\begin{equation}
  v_p=
  \frac{E_\psi(\mathcal R(c_p))}
       {\left\|E_\psi(\mathcal R(c_p))\right\|_2}
  \in\mathbb R^k,
  \qquad
  V_e=
  \begin{bmatrix}
    v_1^\top\\
    \vdots\\
    v_{P_e}^\top
  \end{bmatrix}
  \in\mathbb R^{P_e\times k}.
  \label{eq:semantic_basis}
\end{equation}
Matrix $V_e$ remains frozen during policy training and deployment.
The unit normalization controls only embedding scale and does not change the field semantics of a relation card.

The cards also determine an action-selection matrix $M_e^a\in\{0,1\}^{P_e\times d_a}$, an observation-selection matrix $M_e^o\in\{0,1\}^{P_e\times d_o}$, and delay bands $\{\mathcal L_p\}_{p=1}^{P_e}$.
Row $p$ selects the action and visible response for relation $p$, while $\mathcal L_p$ gives the physical-time range in which a response may be collected after the action.
The effect direction remains in the full relation card and its semantic direction $v_p$, while the selection matrices perform only field routing.
Together, these offline objects define the basic action-relation coordinates that remain fixed across episodes.
One action may correspond to several cards, and one observation may receive several action paths, but duplicate cards should be merged before freezing.
Every action coordinate must be covered by at least one relation, and the final relation order must remain identical across caches, matrices, and training runs.
Thus, $\mathcal K_e$, $V_e$, the two selection matrices, and the delay bands form one versioned offline artifact rather than a stochastic prompt result regenerated between episodes.

\subsection{History-Aligned Relation Strength}\label{sec:relation_strength}

A numerical history encoder first produces the state representation
\begin{equation}
  h_t=E_\theta(\mathcal H_t)\in\mathbb R^{d_h}.
  \label{eq:history_encoder}
\end{equation}
This representation summarizes the overall history needed by the policy, while the relation-strength module explicitly preserves action--response events aligned with documented relations.

To handle irregular sampling, set $\tau_{t-L}=0$ and $\tau_{i+1}=\tau_i+\Delta t_i$, and let $T_t=\tau_t-\tau_{t-L}$ denote the window duration.
Let $M_{e,p}^a$ and $M_{e,p}^o$ denote row $p$ of the two selection matrices.
The valid action--response index set for relation $p$ is
\begin{equation}
  \mathcal I_{t,p}
  =\left\{(i,j):t-L\le i<j\le t,\;
  \tau_j-\tau_i\in\mathcal L_p\right\}.
  \label{eq:index_set}
\end{equation}
This set permits only responses that are already visible at the current instant and satisfy the documented delay.

For every $(i,j)\in\mathcal I_{t,p}$, define the relation-aligned event
\begin{equation}
  z_{t,ij}^{(p)}
  =\left[
  M_{e,p}^a a_i,\;
  M_{e,p}^o(o_j-o_i),\;
  \frac{\tau_j-\tau_i}{T_t},\;
  \frac{\tau_t-\tau_j}{T_t}
  \right]
  \in\mathbb R^4.
  \label{eq:event}
\end{equation}
The first two components are the corresponding historical action amplitude and visible response change, and the last two are normalized response delay and response age.
An event describes observed action--response alignment and is not interpreted as an identified causal effect.
Using change $o_j-o_i$ rather than an absolute observation makes the event focus on the local response after an action, while $h_t$ retains responsibility for modeling the overall operating condition.
Explicit physical time rather than index distance allows one documented delay to be used with irregular sampling.
Response age distinguishes newly observed evidence from older evidence near the start of the window, while the shared map learns how to attenuate or retain this information.

One shared numerical map $\Phi_\xi:\mathbb R^4\rightarrow\mathbb R^{d_r}$ encodes events from every relation, and mean aggregation forms the relation-wise representation:
\begin{equation}
  R_{t,p}
  =\underset{(i,j)\in\mathcal I_{t,p}}{\operatorname{Mean}}
  \Phi_\xi\!\left(z_{t,ij}^{(p)}\right)
  \in\mathbb R^{d_r},
  \qquad
  \operatorname{Mean}_{\emptyset}=0.
  \label{eq:relation_event}
\end{equation}
Using a shared map avoids assigning an independent event network to every documented relation; relation differences are expressed by field selection, delay, and the semantic basis.
Stacking all relation representations gives
\begin{equation}
  R_t=
  \begin{bmatrix}
    R_{t,1}^\top\\
    \vdots\\
    R_{t,P_e}^\top
  \end{bmatrix}
  =\mathcal T_{e,\xi}\!\left(
  \mathcal H_t;M_e^a,M_e^o,\{\mathcal L_p\}
  \right)
  \in\mathbb R^{P_e\times d_r}.
  \label{eq:relation_matrix}
\end{equation}

Numerical events and frozen semantic directions are projected into the same $q$-dimensional matching space.
The relation-wise matching score is
\begin{equation}
  \ell_t
  =\operatorname{diag}\!\left[
  (R_tW_r)(V_eW_v)^\top
  \right]
  \in\mathbb R^{P_e},
  \label{eq:matching}
\end{equation}
where $W_r\in\mathbb R^{d_r\times q}$ and $W_v\in\mathbb R^{k\times q}$ are trainable projections.
The diagonal operator matches numerical events for relation $p$ only with documented relation $p$, avoiding unconstrained cross-relation mixing.

The matching score is converted into relation gains through the centered map
\begin{equation}
  g_t=2\sigma(\ell_t)\in(0,2)^{P_e},
  \label{eq:gain}
\end{equation}
where $\sigma$ is the logistic function.
$g_{t,p}=1$ denotes neutral documented strength, while $g_{t,p}<1$ and $g_{t,p}>1$ denote suppression and enhancement of relation $p$ by recent history.
When a relation has no valid historical event, $R_{t,p}=0$, so $\ell_{t,p}=0$ and $g_{t,p}=1$.
This default preserves the basic documented relation rather than deleting an action path because recent evidence is unavailable.
The neutral fallback also avoids equating ``no valid event has been observed'' with ``the relation is currently ineffective.''
The former is missing evidence, whereas the latter requires recent events that support a smaller gain.
This distinction prevents cold-start windows, long-delay relations, and temporary missing observations from automatically creating zeroed action channels.

\subsection{State-Conditioned Action-Effect Field}\label{sec:effect_field}

For any action $a\in\mathcal A_e$ currently evaluated by the critic, the relation-level action amplitude is
\begin{equation}
  u_e(a)=M_e^a a\in\mathbb R^{P_e}.
  \label{eq:action_amplitude}
\end{equation}
The same action coordinate may appear in several relations because one manipulated variable may affect several process observations or delay paths.
Relation gains and the current action are jointly activated by the same bounded nonlinearity:
\begin{equation}
  \eta_t(a)
  =\tanh\!\left(g_t\odot u_e(a)\right)
  \in(-1,1)^{P_e}.
  \label{eq:activation}
\end{equation}
This operation preserves the action sign, allows relation gains to continuously modulate action sensitivity, and bounds the relation amplitude produced by large normalized actions.

The LCAE state-conditioned action-effect field is defined as
\begin{equation}
  \boxed{
  f_t(a)
  =\eta_t(a)^\top V_e
  =\tanh\!\left(g_t\odot M_e^a a\right)^\top V_e
  \in\mathbb R^k.}
  \label{eq:effect_field}
\end{equation}
This field combines three sources with distinct roles: $V_e$ gives relation identity, $g_t$ gives history-conditioned strength, and $M_e^a a$ gives the amplitude of the currently evaluated action.
The same action therefore usually has different effect fields under different histories, while different actions remain distinct under the same history.
This is not ordinary concatenation of a document embedding, history vector, and action vector because the three jointly determine the critic's action input through a relation-wise nonlinearity.

The local sensitivity of the gain-modulated activation to relation-level action amplitude is
\begin{equation}
  \frac{\partial\eta_{t,p}(a)}{\partial u_{e,p}(a)}
  =g_{t,p}\left(1-\eta_{t,p}(a)^2\right)>0.
  \label{eq:local_sensitivity}
\end{equation}
Because this derivative is strictly positive, the gain does not reverse the action direction in decision coordinates.
It changes local sensitivity, while $\tanh$ gradually reduces sensitivity in large-action regions.
Relation direction is retained by the frozen semantic basis, and numerical history modulates only its current strength.

To prevent the effect field from collapsing valid action directions, its Jacobian with respect to the original decision action should have full column rank:
\begin{align}
  J_t(a)
  &=\frac{\partial f_t(a)}{\partial a} \nonumber\\
  &=V_e^\top
  \operatorname{Diag}\!\left[
  g_t\odot\left(1-\eta_t(a)^2\right)
  \right]M_e^a,
  \qquad
  \operatorname{rank}J_t(a)=d_a.
  \label{eq:jacobian}
\end{align}
Because every diagonal term is strictly positive, a sufficient structural condition is $\operatorname{rank}(M_e^a)=d_a$ and $\operatorname{rank}(V_e)=P_e$.
The first condition requires every valid action direction to be covered by an independent set of relations, and the second requires the semantic relation directions to have no row-rank degeneration.
These conditions should be checked when building the relation basis; if they fail, relation cards must be corrected, merged, or added rather than allowing action-direction loss to remain hidden during training.

\subsection{Decision Backbone and Training Objective}\label{sec:training}

LCAE defines twin critics from the history representation and action-effect field:
\begin{equation}
  Q_m(\mathcal H_t,a;V_e)
  =C_{\phi_m}\!\left(h_t,f_t(a)\right),
  \qquad m\in\{1,2\}.
  \label{eq:critics}
\end{equation}
The critic receives no raw-action bypass, so its action evaluation must pass through the relation field.
The environment-specific actor is
\begin{equation}
  a_t\sim\pi_\omega(\cdot\mid\mathcal H_t)
  :=\pi_\omega(\cdot\mid h_t,g_t).
  \label{eq:actor}
\end{equation}
The actor uses the overall history representation $h_t$ and relation-wise gains $g_t$ but does not receive an arbitrary concatenation of frozen semantic vectors.
Relation identity enters policy learning through the gain indices and the critic's action field.

For replay transition $(\mathcal H_t,a_t,r_t,\Delta t_t,\mathcal H_{t+1},d_t)$, where $d_t$ is the termination indicator, the maximum-entropy Bellman target is
\begin{align}
  y_t
  =r_t
  &+(1-d_t)\exp(-\lambda\Delta t_t) \nonumber\\
  &\cdot\mathbb E_{a'\sim\pi_\omega(\cdot\mid\mathcal H_{t+1})}
  \left[
  \min_m\bar Q_m(\mathcal H_{t+1},a';V_e)
  -\alpha\log\pi_\omega(a'\mid\mathcal H_{t+1})
  \right],
  \label{eq:bellman_target}
\end{align}
where $\bar Q_m$ denotes a target critic.
The temporal discount directly uses the realized $\Delta t_t$ in the transition, keeping different action durations consistent in the value target.

The twin critics are trained by minimizing
\begin{equation}
  \mathcal L_Q
  =\mathbb E\!\left[
  \sum_{m=1}^{2}
  \left(Q_m(\mathcal H_t,a_t;V_e)-y_t\right)^2
  \right].
  \label{eq:critic_loss}
\end{equation}
The reparameterized actor objective is
\begin{equation}
  \mathcal L_\pi
  =\mathbb E_{a\sim\pi_\omega(\cdot\mid\mathcal H_t)}\!\left[
  \alpha\log\pi_\omega(a\mid\mathcal H_t)
  -\min_mQ_m(\mathcal H_t,a;V_e)
  \right].
  \label{eq:actor_loss}
\end{equation}
The actor gradient passes through the critic that depends on $f_t(a)$, so the relation basis does more than provide extra context: it directly shapes action updates.
At the same time, the actor explicitly receives $g_t$, allowing it to use the same history-conditioned relation strengths before generating an action.

The parameter sets $E_\theta$, $\Phi_\xi$, $W_r$, $W_v$, $C_{\phi_1}$, $C_{\phi_2}$, and $\pi_\omega$ are trained by the decision losses above.
$F_{\mathrm{LLM}}$, $E_\psi$, $V_e$, $M_e^a$, $M_e^o$, and $\{\mathcal L_p\}$ remain frozen during policy training.
This separation prevents the policy loss from rewriting offline document relations into unauditable online semantic objects.
It also allows the same relation artifacts to be reused across random seeds and matched baselines under a fixed experimental protocol.

To separate method gains from extra model capacity, a raw-action baseline should share the history encoder, actor and critic widths, optimizer, replay data, environment-interaction budget, and seed path.
Because LCAE adds relation-event and projection parameters, matching hidden widths alone does not constitute strict capacity matching.
A parameter-count-matched numerical baseline without semantic content is therefore also required, together with ablations that keep the architecture fixed while shuffling or removing relation content.
These comparisons are not part of the method definition itself, but they are necessary interfaces for testing whether the action-effect field actually uses documented relations.

\subsection{Deployment and Claim Boundaries}\label{sec:deployment}

At deployment, the system first computes $h_t$, $R_t$, and $g_t$ from visible numerical history, after which the actor generates decision action $a_t$.
The action is mapped back to physical coordinates through $(\mathcal N_e^a)^{-1}$ and sent to the environment.
The critic, LLM, and embedding model do not participate in per-step online execution.
Online computation contains only history encoding, relation-event aggregation, fixed matrix operations, and one actor forward pass.

Different industrial environments share the relation-card schema, action-effect-field definition, and training interface, but they do not share one controller.
Each environment retains its own observation space, action space, reward, coordinate transforms, history length, delay bands, encoder, actor, and critic parameters.
This work therefore does not claim direct policy transfer across environments without retraining.

Relation gain $g_t$ is a history-conditioned action-evaluation quantity, not a physical gain or a strict causal coefficient.
Unobserved disturbances, closed-loop feedback, and sensor errors may all affect observation changes after an action.
LCAE does not perform fault diagnosis, measurement correction, safety shielding, or causal identification; these capabilities require separate evidence and dedicated mechanisms.
The method also assumes that fixed documents are sufficiently accurate and that relation cards cover every valid action direction.
When documents are outdated, relations conflict, or action coverage is insufficient, the corresponding relation artifacts should be withheld and revalidated before deployment rather than left for the online policy to repair automatically.
The deployed system should also bind an offline-artifact hash to the policy checkpoint so that the controller cannot load a mismatched relation order or coordinate definition.
If a history window is incomplete, the system may use the defined neutral fallback; if the schema or coordinate transform does not match, execution must be rejected rather than silently degraded.
This distinction separates recoverable data sparsity from configuration errors that would change action meaning.

\section{Conclusion}\label{sec:conclusion}

This article introduced LCAE, which replaces anonymous continuous actions in industrial actor--critic learning with a state-conditioned action-effect field.
Fixed industrial documents define action--observation--direction--delay relations through an offline LLM, recent numerical history modulates relation strength, and the current action provides the relation amplitude being evaluated.
The same relation gains inform action generation, and the same semantic basis supports long-term value learning, so document knowledge enters policy optimization while the deployed loop remains free of LLM calls.

The central claim is a testable decision bias rather than an unmeasured performance advantage.
When documented relations are correct and recent action--response history reflects contextual strength, LCAE should improve policy learning under limited interaction and changing action effectiveness; matched baselines and shuffled-relation ablations can directly falsify this hypothesis.
Relation gains should not be interpreted as identifiable causal coefficients, and different environments still require their own controllers, rewards, and training processes.
Subsequent validation should first examine relation-card correctness, action coverage, non-semantic baselines with matched information and capacity, and closed-loop outcomes after documented relations are removed or shuffled.

\bibliographystyle{IEEEtran}
\bibliography{references}

\end{document}